\documentclass[twocolumn]{aastex701}

\begin{document}






\title{A Two-month, Galaxy-targeted NIR Follow-up of the Sub-solar-mass Gravitational Wave Candidate S251112cm: Probing Electromagnetic Counterpart Scenarios}

\author[orcid=0000-0002-6639-6533,gname='Gregory', sname='Paek']{Gregory S. H. Paek}
\affiliation{Institute for Astronomy, University of Hawaii, 2680 Woodlawn Drive, Honolulu, HI 96822, USA}
\email{gregorypaek94@gmail.com}
\author[orcid=0000-0002-5115-6377,gname='Felipe', sname='Olivares~E.']{Felipe Olivares~E.}
\affiliation{UKIRT Observatory, Institute for Astronomy, 640 N.\ A'ohoku Place, University Park, Hilo, Hawai'i 96720, USA}
\email{foe@hawaii.edu}
%
\author[0000-0003-3953-9532]{Willem~B.~Hoogendam}
\altaffiliation{NSF Graduate Research Fellow}
\affiliation{Institute for Astronomy, University of Hawai'i at Manoa, 2680 Woodlawn Dr., Hawai'i, HI 96822, USA }
\email{willemh@hawaii.edu}


\begin{abstract}

S251112cm is a nearby ($93 \pm 27$~Mpc) compact-binary merger candidate with a high inferred probability that at least one component is below $1\,M_\odot$.
This event provides a rare test case for electromagnetic follow-up of a theoretically predicted but unconfirmed sub-solar-mass merger candidate.
Motivated by scenarios predicting NIR-bright emission over weeks to months, we carried out a deep, galaxy-targeted UKIRT $J$-band campaign over approximately two months, prioritizing host candidates by a ranking that combines the three-dimensional localization probability with WISE W1 luminosity as a stellar-mass proxy.
Of $\sim9000$ candidate hosts, we monitored 59, corresponding to 7.7\% of the cumulative weighted host prior.
Difference imaging reveals no convincing transient down to typical depths of $J\sim22$--23~mag (AB).
At this distance, both a Type~Ic-BL supernova template and an AT\,2017gfo-like KN template---including time-shifted variants up to $\Delta t \sim 60$~days---would have remained above our median depth for multiple epochs within the monitored hosts.
Comparison with a canonical KN model grid, summarized through conditional weighted consistent-fraction maps, shows that bright configurations with high dynamical and wind-ejecta masses are strongly disfavored, while less luminous configurations remain largely permissible.
While these constraints are host-limited and do not constitute an event-wide exclusion, this campaign provides the first empirical case study of long-baseline, deep NIR follow-up for a sub-solar-mass merger GW class and demonstrates a host-weighted framework applicable to future galaxy-targeted campaigns.

\end{abstract}

\keywords{Gravitational wave astronomy (675), Gravitational wave sources (677), Gravitational waves (678), Near infrared astronomy (1093), Supernovae (1668), Time series analysis (1916), Time domain astronomy (2109),  Stellar mergers (2157)}


\section{Introduction}\label{sec:intro}

The gravitational-wave (GW) signal from the binary neutron star (BNS) merger GW170817 and its electromagnetic (EM) counterpart, AT\,2017gfo, established the foundation of GW multi-messenger astronomy by identifying kilonova (KN) emission as the characteristic optical/NIR diagnostic signature \citep{2017Natur.551...80K, 2017Natur.551...85A, 2017Sci...358.1556C, 2017Sci...358.1570D, 2017Sci...358.1574S, 2018ApJ...855..103P}.
Since then, the LIGO--Virgo network, later joined by KAGRA to form the LIGO--Virgo--KAGRA (LVK) network, has progressed through Observing Runs O1--O4 with steadily improving sensitivity, diversifying the compact-binary merger catalog \citep{2021PhRvX..11b1053A, 2023PhRvX..13d1039A,2025arXiv250818082T}.

A critical development in the fourth Observing Run (O4) has been the emergence of sub-solar mass (SSM) candidates.
In these events, at least one component carries $m \lesssim 1\,M_\odot$.
These objects cannot be produced through standard stellar evolution and their EM counterpart properties remain largely unconstrained.

The physical nature of SSM compact objects remains uncertain, with proposed formation channels including primordial black holes (PBH; \citealt{2024PhRvD.109l4063C}), exotic compact objects stabilized by non-standard nuclear equations of state \citep{2025PhRvD.111h3538C}, and sub-solar-mass neutron stars (ssNSs) produced by disk fragmentation in collapsar environments \citep{2024ApJ...971L..34M, 2025ApJ...991L..22C}.
Because the formation channel determines the merger environment, the expected EM counterpart differs substantially across scenarios.

In the collapsar disk-fragmentation channel, r-process ejecta from an ssNS merger embedded within the overlying SN envelope produce a kilonova-within-a-supernova (KN-in-SN): a transient resembling a stripped-envelope SN (e.g., Type Ic-BL SN) but with a distinctive late-time NIR excess emerging over weeks to months after the explosion, driven by the high opacity of r-process material \citep{2024ApJ...971L..34M, 2025ApJ...991L..22C}.
In more massive collapsar disk environments, the larger r-process ejecta mass ($\gtrsim 5$--$10\,M_\odot$) yields a super-kilonova (super-KN): a redder, longer-lived transient in which lanthanide-rich opacity and line-blanketing shift the spectral energy distribution into the NIR ($\gtrsim 10{,}000$\,\AA), with a peak luminosity of $10^{41}$--$10^{43}$\,erg\,s$^{-1}$ reached on timescales of 10--70\,days and emission sustained over a month or more \citep{2019Natur.569..241S, 2022ApJ...941..100S}.
Unlike the KN-in-SN scenario, the super-KN model does not itself explain the origin of the SSM GW signal. We consider it only as a phenomenological benchmark for bright, long-lived NIR emission that can be tested by our observing strategy.

Even under a standard KN interpretation, the NIR emission remains above optical flux at phases $\gtrsim 5$\,days as dynamical ejecta transition to lanthanide-rich conditions \citep{2017Natur.551...80K, 2020LRR....23....1M} and rapid post-peak fading of $\Delta m \gtrsim 0.1$\,mag\,day$^{-1}$.
Across all these scenarios, the defining observational feature is a late-time NIR excess on timescales of weeks to months.
Such emission is inaccessible to optical-only campaigns and invisible to searches that terminate within days of the GW trigger.
An EM counterpart detection, or stringent upper limits tied to these model predictions, provides the primary empirical handle on the progenitor channel.

The first notable SSM candidate in O4, S250818k (FAR $\sim\!2.1\,\mathrm{yr}^{-1}$), prompted multi-facility follow-up during which the Type~IIb SN (SN\,2025ulz) was identified within the localization volume as a possible counterpart; however, subsequent monitoring indicated that the association is likely coincidental \citep{2026ApJ..1001L..20H,2025ApJ...995L..27G,2025ApJ...994L..45F,2025ApJ...995L..59K,2025arXiv251024620H}.
Notably, the follow-up efforts for S250818k were concentrated in optical bands and confined to timescales of days post-trigger, a strategy well-suited to canonical KN searches but incapable of constraining emission predicted for super-KNe and KN-in-SN scenarios.

The second notable SSM candidate, S251112cm, was identified on 2025 November 12 at 15:18:45 UTC by the MBTA SSM low-latency pipeline \citep{2025GCN.42650....1L, 2025GCN.42653....1C}, with a FAR of $\sim\!5.1 \times 10^{-9}$\,Hz ($\sim\!1$ per 6.2\,yr), a luminosity distance of $93 \pm 27$\,Mpc, and a posterior probability $\mathrm{HasSSM} = 100\%$ \citep{2025GCN.42693....1C}, making it the most significant SSM GW candidate reported to date.

The event triggered an extensive community response, with optical and NIR follow-up reported by GOTO \citep{2025GCN.42658....1A}, BlackGEM/MeerLICHT \citep{2025GCN.42663....1G}, ATLAS \citep{2025GCN.42666....1G, 2025GCN.42682....1S}, TROVE \citep{2025GCN.42675....1F}, ZTF \citep{2025GCN.42677....1A}, DECam/GW-MMADS \citep{2025GCN.42691....1H}, GRANDMA \citep{2025GCN.42698....1B}, Rubin/LSST \citep{2025GCN.42707....1M, 2025GCN.43257....1A}, WFST \citep{2025GCN.42722....1L}, SOAR \citep{2025GCN.42724....1S, 2025GCN.42796....1B}, SVOM/VT \citep{2025GCN.42777....1M}, Kinder \citep{2025GCN.42825....1G}, and GECKO/KMTNet \citep{2025GCN.42867....1J}.
In particular, a wide-field optical campaign using Pan-STARRS and ATLAS \citep{smith2026inprep} covers most of the localization area when combined with Rubin coverage, but reports no convincing counterpart.
Despite this breadth, these campaigns shared the same structural limitation as prior SSM follow-up: observations were predominantly optical and/or confined to short baselines of days to one week post-trigger, leaving a potential late-time NIR emission phase unconstrained.

In this work, we present a deep, galaxy-targeted UKIRT $J$-band monitoring campaign for the highest-probability hosts within the S251112cm localization volume, spanning two months post-trigger at a several-day cadence.
The campaign is designed to probe both the predicted peak phase of super-KNe ($t_{\rm peak} \sim 10$--70~days) and the late-time NIR excess characteristic of KN-in-SN scenarios.
We then evaluate the resulting upper limits against benchmark KN-like and SN-like templates, framed within a host-weighted framework that quantifies which regions of counterpart parameter space are excluded under the conditional assumption that the source lies in the monitored host subset.


\section{Observations and Data Processing} \label{sec:observation}



\subsection{Galaxy-targeted Observing Strategy}

\begin{figure*}[t]
  \centering
  \includegraphics[width=\linewidth]{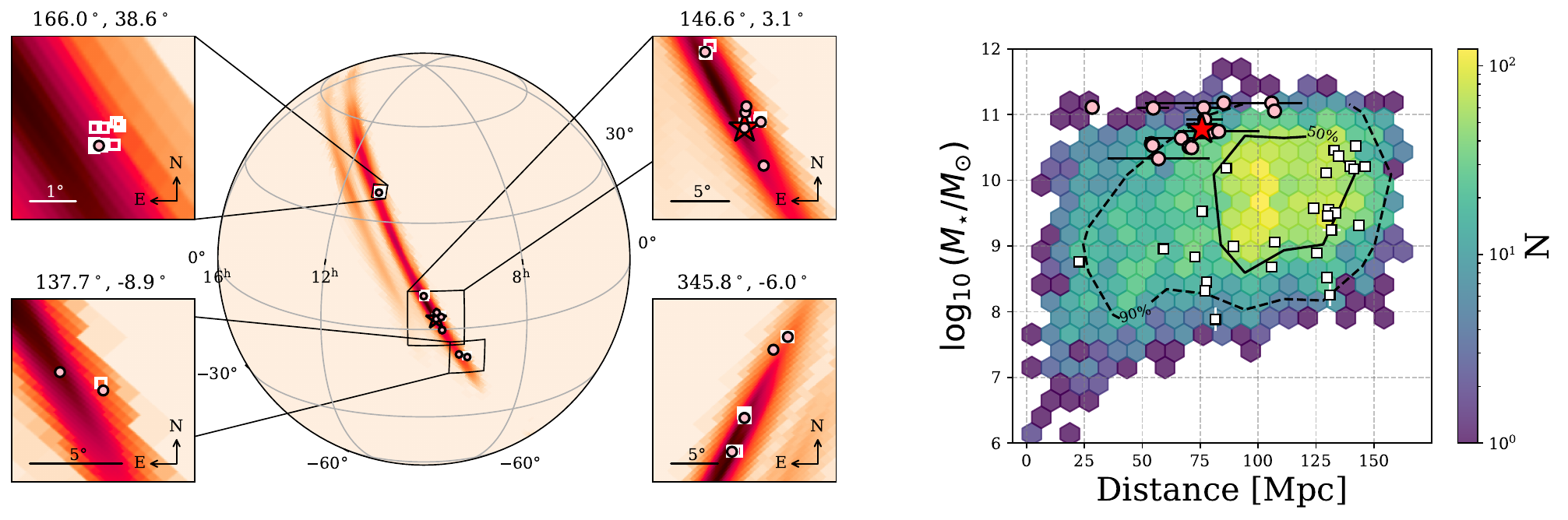}
  \caption{
  \textit{Left}: GW localization probability map of S251112cm (Bilby pipeline; \citealt{2025GCN.42690....1L}) in an orthographic globe projection.
Four insets show zoomed-in regions where candidate host galaxies are concentrated.
The upper-right inset is centered on the highest-ranked host candidate (IC\,0564), while the center coordinates of the remaining three insets are indicated in their respective titles.
  The red star marks the highest-ranked host candidate, pink circles mark 16 monitored hosts within the top 50 of the full catalog ranking, and square markers indicate the remaining 42 monitored hosts; some symbols overlap at the scale shown.
  The Rank IDs in Table~\ref{tab:galaxy} encode positions in the full catalog ranking and can be used to identify the plotted hosts.
  \textit{Right}: Distribution of the matched 59 galaxies with the localization volume of S251112cm with NED gravitational-wave catalog in luminosity distance versus stellar mass, with the same marker indicators as in the left figure.
  The hexbin color scale indicates the number density of catalog 9,047 galaxies within localization volume of S251112cm.
  Black solid and dashed contours enclose 50\% and 90\% of the catalog, respectively.
  }
  \label{fig:gw_loc}
\end{figure*}

The sky localization of S251112cm is too broad for an efficient blind wide-area $J$-band search with UKIRT.
We therefore adopted a galaxy-targeted strategy and re-optimized the target list as the GW localization evolved.
Host candidates were drawn from the NED gravitational-wave galaxy catalog \citep{2023ApJS..268...14C} \footnote{Initial localization: \url{https://ned.ipac.caltech.edu/uri/NED::GWFglist/fits/S251112cm/3$0} \citep{2025GCN.42653....1C}.
Updated localization: \url{https://ned.ipac.caltech.edu/uri/NED::GWFglist/fits/S251112cm/4$0} \citep{2025GCN.42693....1C}.}.
The initial (1220~deg$^2$ at 90\% credible region) and updated (1681~deg$^2$) localization volumes contain 7731 and 9047 unique galaxies from the catalog, respectively (Figure~\ref{fig:gw_loc}).

For each galaxy $i$, we assigned a priority score

\begin{equation}
S_i \propto P_{{\rm 3D},i}\times L_{{\rm W1},i},
\end{equation}

where $P_{{\rm 3D},i}$ is the localization probability density at the galaxy position and distance, and $L_{{\rm W1},i}$ is the WISE W1 luminosity proxy for stellar mass.
Given the unknown host-galaxy preferences of SSMs, we adopt the WISE W1 luminosity as a proxy for stellar mass, following the prescription that higher stellar-mass galaxies are preferred host candidates for BNS and NSBH mergers, based on host-galaxy preference empirical relations and simulations \citep{2019MNRAS.487....2M, 2019MNRAS.487.1675A, 2020MNRAS.491.3419A, 2020MNRAS.495.1841A}.

\begin{figure}[t]
  \centering
  \includegraphics[width=\linewidth]{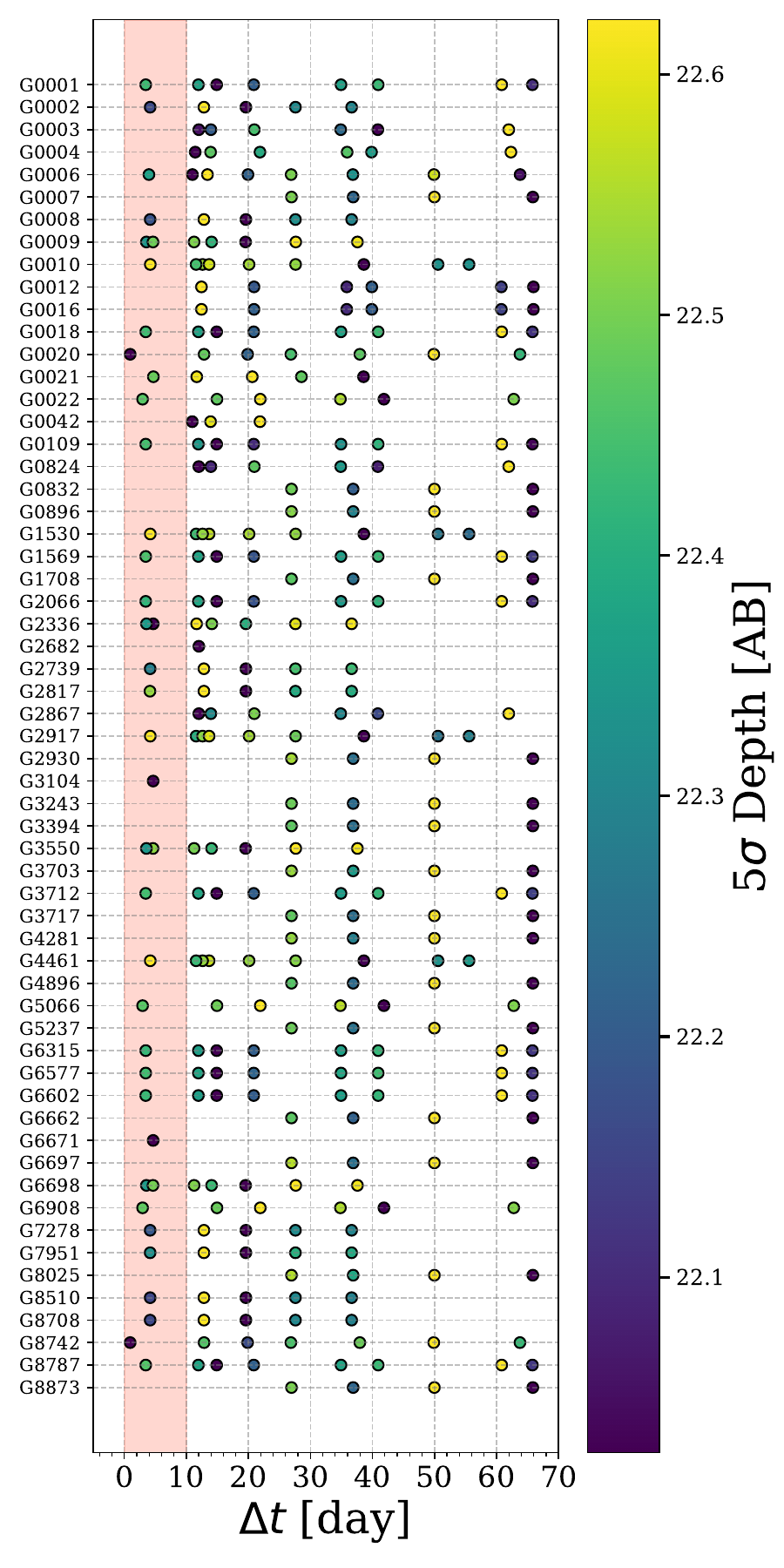}
  \caption{Summary of time-series monitoring for the monitored hosts within the GW localization volume.
  Marker colors encode days after the GW trigger ($\Delta t$) and 5$\sigma$ depth.
  The red shaded region marks the first 0--10 days after the GW trigger, indicating the early-phase observations.
  Each host nickname corresponds to Table~\ref{tab:galaxy}.}
  \label{fig:obs_summary}
\end{figure}

Monitoring began after the initial alert \citep{2025GCN.42650....1L}, was re-optimized after the updated localization \citep{2025GCN.42690....1L}, and continued for approximately two months with the highest cadence in the early phase and lower cadence at later times (Figure~\ref{fig:obs_summary}).

Observations were carried out with the Wide-Field Camera (WFCAM) on UKIRT between 2025 November 13 and 2026 January 17 (UT) under program U/25B/H03.
All science frames were obtained in the $J$ band ($\lambda_{\rm eff} \approx 1.25\,\mu$m) with 10\,s individual exposures.
Each pointing followed a 5-point jitter dither pattern, and a nominal observing block consisted of 8 dither sequences yielding 40 frames and 400\,s of on-sky integration per visit.
A subset of early-run nights (2025 November 13--17) used a shallower configuration of 4 dither sequences (200\,s), and several later visits were truncated by weather or queue constraints.
The cumulative on-sky integration per target ranged from 200\,s to approximately 4000\,s across the full campaign, with the upper end reflecting targets that fell within the WFCAM field of view of multiple pointings.

\begin{deluxetable}{cccc}
  \tablecaption{Observation log for each monitored host in this work.
  We use the Rank ID defined in Table~\ref{tab:galaxy}.
  Here $\Delta t \equiv t_{\rm obs} - t_0$, where $t_0$ is the GW trigger time corresponding to 2025-11-12 15:18:45.362 UT.}
  \label{tab:obslog}
  \tablewidth{0pt}
  \tablehead{
    \colhead{Rank ID} & \colhead{$\Delta t$ (day)} & \colhead{$5\sigma$ Depth} & \colhead{Seeing (arcsec)}
  }
  \startdata
    G0001 & 3.462 & 22.508 & 1.389 \\
    G0001 & 11.945 & 22.400 & 1.491 \\
    G0001 & 14.894 & 21.844 & 1.082 \\
    G0001 & 20.863 & 22.133 & 1.471 \\
    G0001 & 34.929 & 22.396 & 1.347 \\
    G0001 & 40.917 & 22.493 & 1.391 \\
    G0001 & 60.824 & 22.804 & 1.296 \\
    G0001 & 65.778 & 21.972 & 1.736 \\
  \enddata
  \tablecomments{Only G0001 is shown here to illustrate the format of the observation log.
  Machine-readable table will be provided.}
\end{deluxetable}

In total, 59 hosts---including those serendipitously captured within the field of view---were observed at least once with repeat $J$-band imaging (Figure~\ref{fig:gw_loc}; see Table~\ref{tab:obslog} for the observation log and Table~\ref{tab:galaxy} in Appendix~\ref{appendix:galaxy_catalog} for the full host catalog).
Due to target visibility, the evolving localization, and the serendipitous inclusion of galaxies within UKIRT/WFCAM pointings, the monitored sample does not exactly correspond to the top 59 galaxies in the final ranking.

\subsection{Data Reduction and Image Subtraction}

Initial processing used the standard UKIRT/WFCAM reduction pipeline products, including dark subtraction, flat-fielding, sky subtraction, and artifact handling.
We then stacked aligned exposures with \texttt{SWarp} \citep{2010ascl.soft10068B} to produce per-epoch science mosaics.

Source detection and photometric measurements were performed with \texttt{Source Extractor} \citep{1996A&AS..117..393B}.
Zero points were derived from point sources in the image matched with 2MASS Point Source Catalog (2MASS PSC; \citealt{2006AJ....131.1163S}), and all photometry was placed on a common AB system using a consistent Vega-to-AB conversion across epochs. For the $J$ band, we adopted

\begin{equation}
m_{\rm AB} = m_{\rm Vega} + 0.91,
\end{equation}

where $m_{\rm AB}$ is the AB magnitude, $m_{\rm Vega}$ is the corresponding Vega magnitude, and 0.91 is the adopted $J$-band Vega-to-AB offset \citep{2007AJ....133..734B}.


Difference image analysis was performed with \texttt{HOTPANTS} \citep{2015ascl.soft04004B}.
For each host galaxy candidate, the reference image was selected as a single epoch simultaneously satisfying a seeing threshold of $<1\farcs4$ and a $5\sigma$ limiting magnitude of $>22.6$\,mag; among epochs meeting both criteria, the deepest image was chosen, with best seeing as a tiebreaker.
If no epoch satisfied both criteria, the deepest available image was adopted as the reference.
Because the chosen reference epoch could in principle contain transient flux and thereby suppress its detection in difference imaging, we additionally performed a brute-force multi-reference cross-check: for each host galaxy candidate, every individual epoch was used in turn as a reference, and the resulting set of difference images was visually inspected for residuals consistent with a fading transient.
No residuals consistent with a fading transient were identified in any host galaxy candidate across the multi-reference set.




\section{Results} \label{sec:result}



\subsection{Follow-up Coverage and Depth}

Table~\ref{tab:obslog} lists the per-epoch observation log, including the elapsed time from the GW trigger and the $5\sigma$ limiting depth for each visit.
The monitored sample is preferentially distributed at the high-stellar-mass end of the candidate host catalog, as expected from the prioritization scheme described in Section~\ref{sec:observation} (right panel of Figure~\ref{fig:gw_loc}).
Relative to the full catalog of $\sim\!9000$ candidate hosts, this monitored subset accounts for 7.7\% of the cumulative $P_{\rm 3D} \times L_{W1}$ weighted host prior.
Typical depth in the final difference-image search is $J \sim 22$--23~mag (AB; Figures~\ref{fig:gw_loc} and \ref{fig:obs_summary}).



Visual inspection for transient candidates was performed in the difference images within 1~arcmin of each host.
We find no convincing transient candidate associated with the monitored host galaxies.

We compiled a sample of 115 unique optical/NIR transients associated with S251112cm from the GCN Circulars cited in Section~\ref{sec:intro}.
We cross-matched these reported locations against the sky footprints of our UKIRT observations, but none coincided with the monitored fields.
In summary, neither visual inspection of difference images around the monitored hosts nor cross-matching against the GCN-reported transient list yields a convincing counterpart candidate associated with our UKIRT monitoring.

\subsection{Model Constraints}

\subsubsection{Benchmark KN-like Constraints}
\label{sec:kn_benchmark}
To assess the information content of the non-detections, we compare per-host $J$-band limits with a benchmark KN model grid from \citet{2021ApJ...918...10W}.
This comparison is phenomenological and is used as a benchmark only. It is not a dedicated EM model for SSM mergers.
In particular, the grid spans ejecta masses $m \in \{0.001, 0.003, 0.01, 0.03, 0.1\}\,M_\odot$ calibrated to BNS merger expectations. 
Whether an SSM merger yields systematically smaller ejecta than this range is not obvious a priori. Numerical-relativity simulations of an asymmetric subsolar binary by \citet{2026arXiv260325102C}, motivated in part by SSM candidates such as S251112cm, find that a highly asymmetric mass ratio, expected when one component is sub-solar, can instead enhance the dynamical ejecta by a factor of $\sim30$ relative to an equal-mass binary of the same total mass, as tidal disruption of the lower-mass component becomes the dominant ejection channel. The mass ratio of S251112cm itself is not well constrained, so this comparison is illustrative rather than event-specific, but it shows that the ejecta mass depends on mass ratio, compactness, and equation of state rather than on total mass alone. We therefore treat the full mass range of the grid as a phenomenological benchmark for SSM events, without assuming a priori that its lower-mass end is the more physically relevant regime.

The canonical KN model grid comprises 900 two-component 2D axisymmetric KN models simulated with the Monte Carlo radiative transfer code \texttt{SuperNu}, incorporating a full suite of lanthanide and fourth-row element opacities.
The two ejecta components are a toroidal low-$Y_e$ component representing tidally driven dynamical ejecta with a robust r-process composition, and a spherical (TS) or lobed (TP; ``peanut-shaped'') high-$Y_e$ component representing wind or shock-driven outflows.
Each component is independently varied over ejecta mass and average velocity $v \in \{0.05, 0.15, 0.3\}\,c$; higher ejecta mass increases luminosity and broadens the light curve, while higher velocity shifts the peak to earlier times and narrows it.
Model light curves are provided over viewing angles from one pole ($0\arcdeg$), through the edge-on orientation ($90\arcdeg$), to the opposite pole ($180\arcdeg$).
The edge-on view is more heavily obscured by the lanthanide-rich low-$Y_e$ component (lanthanide curtaining), suppressing optical flux and enhancing NIR emission relative to either axial view.

\begin{figure}[t]
  \centering
  \includegraphics[width=\linewidth]{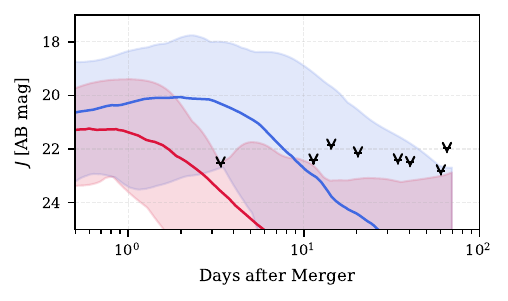}
  \caption{Example benchmark KN-like constraint for the monitored host IC\,0564 (G0001), derived from the model-comparison procedure described in Section~\ref{sec:result}.
  Blue curves denote model realizations brighter than the $5\sigma$ upper limits and therefore constrained by the non-detections; the blue shaded region spans the minimum--maximum range of these constrained models, and the blue line marks their median.
  Red curves denote model realizations fainter than the upper limits and therefore not constrained by the data.}
  \label{fig:kn_constraint_example}
\end{figure}

The grid is scaled to the distance of each observed host galaxy, and the resulting $J$-band model light curves are compared with the per-host upper limits at each observed epoch to identify which combinations of ejecta mass, velocity, and viewing angle are excluded by the non-detections.
Figure~\ref{fig:kn_constraint_example} shows an example for the highest-ranked host (IC0564).

\subsubsection{Conditionally Weighted Consistent Fraction}\label{subsec:condition}

To summarize the benchmark constraints across the monitored hosts, we define explicit host weights using the same ranking metric adopted for target selection,
\begin{equation}
w_i = \frac{(P_{\rm 3D} \times L_{\rm W1})_i}{\sum_{j \in \mathrm{all\ host}} (P_{\rm 3D} \times L_{\rm W1})_j},
\label{eq:host_weight}
\end{equation}
where $w_i$ is the absolute contribution of host galaxy $i$ to the weighted host prior. The numerator is the host-specific ranking metric, and the denominator is taken over the full host catalog rather than only the observed subset.

For each host and each KN parameter bin $p$, we define the host-level consistent fraction as
\begin{equation}
C_i(p) = \frac{N_{\mathrm{cons},i}(p)}{N_{\mathrm{total},i}(p)},
\label{eq:host_survive}
\end{equation}
where $p$ denotes a KN parameter bin, $N_{\mathrm{total},i}(p)$ is the total number of model realizations in bin $p$ for host $i$, and $N_{\mathrm{cons},i}(p)$ is the number of those models whose distance-scaled light curves remain fainter than the observational limits at all valid epochs and are therefore consistent with the non-detections. Thus, $C_i(p)=1$ indicates that all models in that bin are permissible for host $i$ given the observational constraints, whereas $C_i(p)=0$ indicates that none are permissible for that host.

We then define the weighted excluded fraction within the monitored host subset,
\begin{equation}
E_{\mathrm{cond}}(p) =
\frac{\sum_{i \in \mathrm{obs}} w_i \left[1 - C_i(p)\right]}{\sum_{i \in \mathrm{obs}} w_i},
\label{eq:excluded_cond}
\end{equation}
which is the weighted fraction of models excluded within the observed subset after normalization by the monitored weighted host prior. The corresponding conditional weighted consistent fraction is
\begin{equation}
S_{\mathrm{cond}}(p) = 1 - E_{\mathrm{cond}}(p).
\label{eq:survive_cond}
\end{equation}

\begin{figure*}[t]
  \centering
  \includegraphics[width=0.96\textwidth]{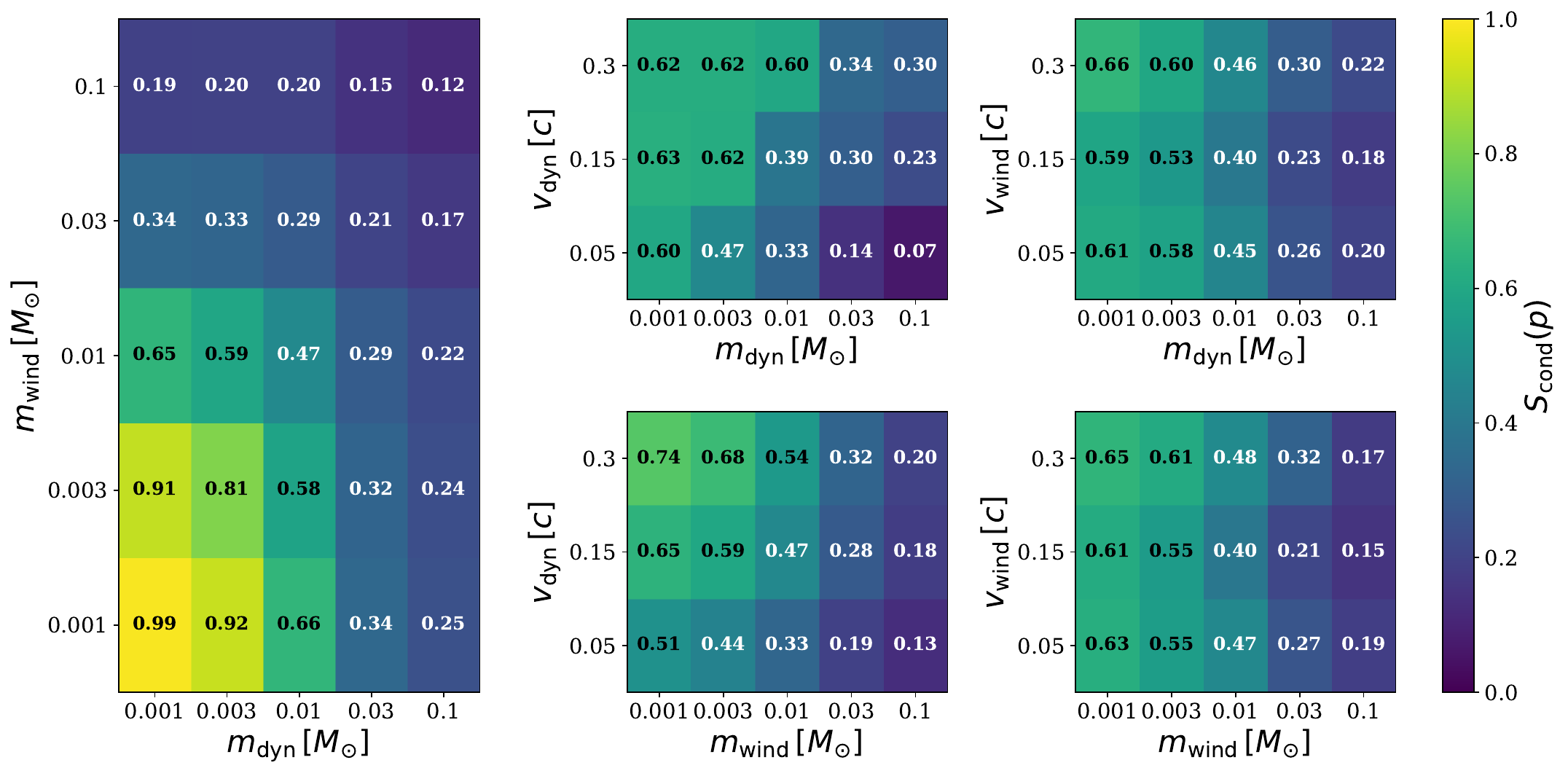}
  \caption{Conditional weighted consistent-fraction map, $S_{\rm cond}(p)$, for the TP case within the monitored host subset. The left panel shows the dynamical-ejecta mass versus wind-ejecta mass plane. The right panels show the mass--velocity combinations. The color scale visualizes the conditional weighted consistent fraction defined by Equations~(\ref{eq:host_weight})--(\ref{eq:survive_cond}). Low consistent fraction indicates strong constraints, whereas high consistent fraction indicates that a larger fraction of the parameter bin remains permissible within the observed host subset.}
  \label{fig:global_tp_scond}
\end{figure*}

\begin{figure*}[t]
  \centering
  \includegraphics[width=0.96\textwidth]{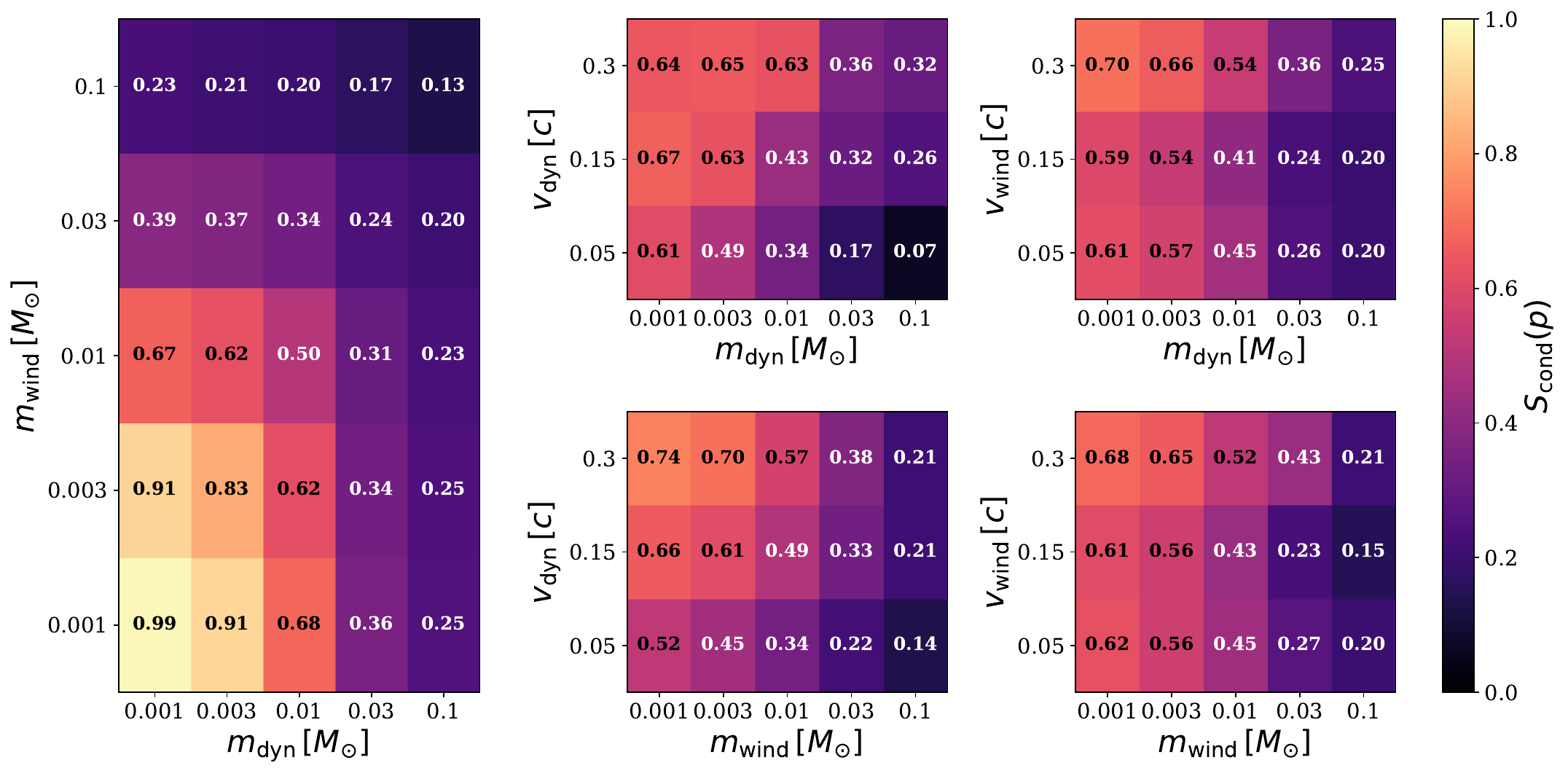}
  \caption{Same as Figure~\ref{fig:global_tp_scond}, but for the TS case.}
  \label{fig:global_ts_scond}
\end{figure*}

Equations~(\ref{eq:host_weight})--(\ref{eq:survive_cond}) therefore define the host weights, the host-level consistent fractions, the weighted excluded fraction, and the conditional weighted consistent fraction used throughout the remainder of this section.
The resulting conditional weighted consistent-fraction maps, shown for the TP and TS cases in Figures~\ref{fig:global_tp_scond} and \ref{fig:global_ts_scond}, respectively, summarize the host-weighted constraints across parameter space. By construction, the plotted quantity is $S_{\rm cond}(p)=1-E_{\rm cond}(p)$: low consistent fraction indicates strong constraints, whereas high consistent fraction indicates that a larger fraction of the model bin remains permissible given the observational constraints.
This is a conditional quantity, not an event-wide posterior probability; it answers the question, ``if the counterpart lies within the monitored weighted host subset, what fraction of models in bin $p$ remains consistent with the non-detections?''
Alternative heatmap representations based on OR-gate combinations without weighting is presented in Appendix~\ref{appendix:alt_heatmaps}.

For both TP and TS morphologies, $S_{\rm cond}$ is highest at low $m_{\rm dyn}$ and low $m_{\rm wind}$ and decreases toward larger ejecta masses (Figures~\ref{fig:global_tp_scond} and \ref{fig:global_ts_scond}).
The same trend is visible in the mass--velocity projections, where combinations producing higher peak luminosities show systematically lower $S_{\rm cond}$.

\begin{figure}[t]
  \centering
  \includegraphics[width=\linewidth]{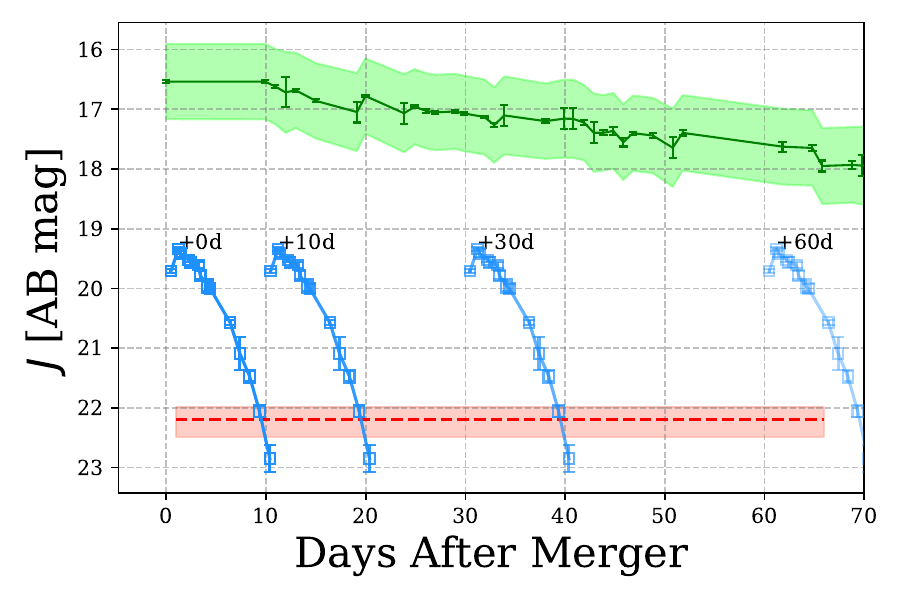}
  \caption{Comparison of empirical transient templates with the UKIRT $J$-band detection threshold, scaled to the luminosity distance of S251112cm ($93 \pm 27$~Mpc).
  \textit{Green line and band}: $J$-band light curve of the Type~Ic-BL supernova SN\,2007I \citep{2014ApJS..213...19B}, with the shaded region reflecting propagated photometric and distance uncertainties.
  \textit{Blue curves}: AT\,2017gfo $J$-band light curve from the ENGRAVE collaboration\footnote{\url{https://www.engrave-eso.org/AT~2017gfo-Data-Release/}} \citep{2022MNRAS.515..631G,2024MNRAS.529.2918G}, shown at time offsets of $\Delta t = 0, 10, 30, 60$~days to illustrate the range of delayed-merger scenarios motivated by collapsar disk-fragmentation models \citep{2024ApJ...971L..34M, 2025ApJ...991L..22C}.
  \textit{Red dashed line and band}: median (dashed) and interquartile range (Q1--Q3; shaded) of the $5\sigma$ $J$-band limiting depths across all monitored epochs and monitored hosts.}
  \label{fig:compare_icbl}
\end{figure}




\subsubsection{Comparison with SN-like and KN-like Empirical Templates}\label{sec:empirical_compare}

As complementary empirical benchmarks, we compare our $J$-band limits with a Type~Ic-BL SN template (SN\,2007I; \citealt{2014ApJS..213...19B}) and the AT\,2017gfo KN $J$-band light curve from the ENGRAVE collaboration \citep{2022MNRAS.515..631G, 2024MNRAS.529.2918G}, both scaled to the luminosity distance of S251112cm (Figure~\ref{fig:compare_icbl}).
The Type~Ic-BL template is motivated by the collapsar disk-fragmentation channel, in which the SN-like outer ejecta enveloping a possible KN-in-SN counterpart are expected to resemble a stripped-envelope SN at early phases \citep{2024ApJ...971L..34M, 2025ApJ...991L..22C}.
We treat the SN-like and KN-like templates in Figure~\ref{fig:compare_icbl} as separate brightness benchmarks rather than as two components to be summed into a single predicted light curve. In the disk-fragmentation picture, the r-process ejecta from the embedded ssNS merger is expected to mix into the surrounding SN-like ejecta rather than emerge above it, so the additional lanthanide and actinide opacity raises the optical depth of the SN ejecta and reddens its observed colors rather than producing an independently visible KN peak \citep{2024ApJ...971L..34M, 2025ApJ...991L..22C}. The true KN-in-SN light curve is therefore better represented by a reddened, NIR-enhanced version of the SN-like template than by the unobscured AT\,2017gfo template alone, and the two templates shown here should be read as bracketing the plausible brightness and timescale of the transient rather than as two independent physical components. We revisit the implications of this reprocessing for our non-detection in Section~\ref{sec:nondetection}.
The AT\,2017gfo template is additionally shown at time offsets of $\Delta t = 10$--60~days to span the range of delayed-merger scenarios predicted by the same models.
Figure~\ref{fig:compare_icbl} shows that at the distance of S251112cm, both the Type~Ic-BL SN template and the AT\,2017gfo-like KN template including time-shifted variants up to $\Delta t \sim 60$~days, lie above our median $5\sigma$ depth at multiple sampled epochs, indicating that such counterparts, if present within the monitored hosts during the observing window, would have been detectable.
Given our $\sim\!4$-day mean cadence and the $\sim\!10$-day window during which an AT\,2017gfo-like KN remains above the median depth at this distance, any such transient peaking within our monitoring baseline, including the $\Delta t = 10$--60~day delayed variants, would have been sampled at multiple detectable epochs.

\begin{figure}[t]
  \centering
  \includegraphics[width=\linewidth]{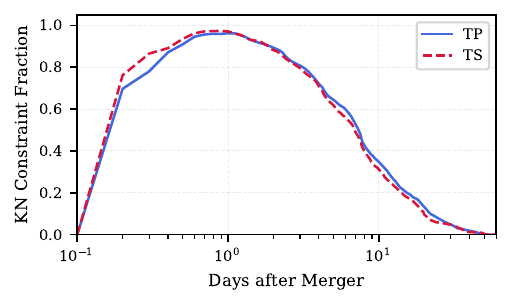}
  \caption{
  Fraction of \citet{2021ApJ...918...10W} KN model grid excluded by a single $J$-band non-detection at $J = 22.2$~mag (median campaign
  depth) scaled to $93$~Mpc, as a function of time after merger.
  Blue solid and red dashed curves correspond to the TP (toroidal$+$peanut) and TS (toroidal$+$spherical) high-$Y_e$ morphology subsets, respectively.
  A model is considered excluded if its predicted $J$-band flux at the given epoch exceeds the detection threshold.
  }
  \label{fig:constraint_fraction}
\end{figure}

\section{Discussion}
\label{sec:discussion}


Our two-month UKIRT $J$-band campaign provides the first long-baseline NIR constraints on EM emission from an SSM GW candidate, reaching $J \sim 22$--23~mag ($M_J \sim -11.8$ to $-12.8$ at 93~Mpc) across the monitored host-galaxy subset, with no transient counterpart detected.

Because the monitored subset accounts for only 7.7\% of the cumulative $P_{\rm 3D} \times L_{\rm W1}$ weighted host prior, all constraints presented below are conditional on the GW source residing within this subset and do not constitute event-wide exclusions.
We therefore first outline the physical scenarios that are consistent with the non-detection (Section~\ref{sec:nondetection}), before quantifying the constraints from individual benchmark comparisons in subsequent subsections.

\subsection{Interpretation of the Non-detection}
\label{sec:nondetection}
The absence of an optical/NIR counterpart in the monitored host subset admits several interpretations that are not mutually exclusive.

First of all, the most straightforward explanation is that the GW source resides outside the monitored 7.7\% of the cumulative $P_{\rm 3D} \times L_{W1}$ weighted host prior.
Although the monitored hosts include most of the top 10 systems ranked by the combined $P_{3D}$ and stellar-mass metric, the total covered prior fraction remains small.

If the GW source does reside within the monitored host subset, our $J$-band imaging reaches approximately 3~mag below the AT~2017gfo-like KN peak at this distance.
This depth rules out canonical KNe and super-KNe within the observed host galaxy candidates, since both classes would exceed our detection threshold at multiple epochs.
A KN-in-SN counterpart could in principle be attenuated or reprocessed by the surrounding SN-like ejecta, even when the intrinsic r-process yield is substantial.
Separately, intervening host-galaxy dust could suppress the observed $J$-band flux.
If the required suppression were attributed to foreground dust, $A_J \approx 3$~mag would correspond to $A_V \approx 10.7$~mag under the standard \citet{1989ApJ...345..245C} extinction law with $R_V = 3.1$, exceeding the typical line-of-sight extinction through giant molecular clouds and embedded star-forming regions \citep{2026ApJ..1002...64C}.
This extinction-law estimate applies only to intervening dust and does not quantify radiative reprocessing within the SN ejecta.
Thus, ordinary host-galaxy extinction alone is unlikely to hide such a counterpart, whereas suppression by the SN ejecta remains model-dependent.

Alternatively, the EM counterpart may be delayed beyond the $\sim\!66$-day baseline of our monitoring campaign.
The KN-in-SN light curve evolves slowly, with its late-time NIR excess not expected to emerge until weeks to months after the explosion (Section~\ref{sec:intro}). Motivated by this timescale, we explore time-shifted AT\,2017gfo templates up to $\Delta t \sim 60$~days in Figure~\ref{fig:compare_icbl} to test the sensitivity of our monitoring baseline to such slowly emerging emission. The delay between the stellar collapse and the merger itself is expected to be considerably shorter. For parameters representative of S251112cm, \citet{2026arXiv260510940H} estimate a maximum of $\sim3.6$~days for the ssNS-ssNS merger and $\lesssim11$~days for the subsequent aftermath chirp with the central BH, so the light curve's own slow evolution, rather than uncertainty in the merger timing, sets the relevant timescale for our template shifts.

The collapsar disk-fragmentation channel predicts a sequence of GW signals from mergers among sub-solar-mass neutron-star fragments, followed by a possible aftermath chirp from the final coalescence with the central BH \citep{2024ApJ...971L..34M, 2025ApJ...991L..22C}.
No precursor or coincident secondary GW signal was reported in association with S251112cm.
Indeed, \citet{2026arXiv260317009V} argue that the expected aftermath chirp should have been detectable at $93$~Mpc by comparison with the more distant GW190814 \citep{2020ApJ...896L..44A}.
This absence weakens the expectation of an EM counterpart arising specifically from this channel, but it does not exclude the scenario because the detectability of the predicted signals depends on the fragment masses and on LVK observing coverage.
In particular, the reported LIGO Hanford and Livingston downtime during the expected aftermath window leaves open the possibility that a secondary signal could have fallen in a coverage gap \citep{2026arXiv260510940H}.
Crucially, this coverage gap underscores the vital role of independent EM follow-up in the multi-messenger framework. 
Because the primary GW network was temporarily blind during this critical window, our long-baseline NIR campaign effectively acts as an empirical "watchdog," providing the only available constraints against the predicted emission with long timescale.

In contrast, an independent consistency test by \citet{2026arXiv260325795R} finds that a primordial black hole (PBH) binary merger interpretation of S251112cm is compatible with existing microlensing and CMB constraints on the PBH abundance.
Under the PBH scenario, no EM counterpart is expected from the merger itself, and a UV/optical flare from accretion onto surrounding gas is possible only if the merger occurred within an AGN disk. No such signal was identified in our monitored host subset or in the broader follow-up campaigns \citep{2026arXiv260510940H, 2026arXiv260317009V, smith2026inprep}.
The PBH interpretation is therefore one of several scenarios consistent with the non-detections, under the assumption the GW source was in one of the monitored galaxies.

Wide-field optical campaigns covering a substantially larger fraction of the localization area report no confirmed counterpart. \citet{2026arXiv260317009V} vetted 248 candidates and found no association with S251112cm after spectroscopic follow-up of the most promising candidates.
\citet{2026arXiv260510940H} covered $\sim\!56\%$ of the localization area within 2.4~hours of the alert using DECam and ZTF.
They excluded 42--92\% of canonical KN models within the surveyed region but likewise reported no confirmed counterpart.
Although a complementary Pan-STARRS and ATLAS campaign with Rubin observations \citep{smith2026inprep} covered the entire localization area, they similarly find no convincing optical counterpart within $\sim\!2$~weeks of the trigger.
Taken together, the optical non-detections imply that any counterpart within the surveyed localization region was either fainter than the contemporaneous optical limits at the observed epochs or rose only after the main optical search window.

In summary, our long-baseline NIR non-detection admits a similarly constrained interpretation.
The most likely interpretation is that the GW source was not associated with one of our monitoring samples.
If present within our samples, its NIR emission was either intrinsically fainter than the benchmark models considered here or delayed beyond the duration of our campaign.
The combined optical and NIR follow-up therefore does not exclude all counterpart scenarios.

\subsection{KN Model Grid Constraints and Observing Epoch Dependence}
\label{sec:kn_grid_discussion}
The pattern visible in Figures~\ref{fig:global_tp_scond} and \ref{fig:global_ts_scond} reflects the sensitivity floor imposed by our median depth of $J \sim 22.2$~mag at $93$~Mpc. Only model realizations whose distance-scaled light curves rise above this threshold during the observed epochs are excluded, while intrinsically faint configurations fall below detection regardless of when they are observed.
As discussed in Section~\ref{sec:kn_benchmark}, the dynamical ejecta mass is sensitive to mass ratio and can be enhanced rather than suppressed for an asymmetric subsolar binary, so the high-mass configurations excluded here cannot be dismissed as unphysical priors on mass grounds alone.
Nonetheless, the lower-mass regime ($\lesssim 0.01\,M_\odot$) remains largely permissible despite our empirical limits, independent of this consideration.
The same asymmetry appears identically in the TP and TS subsets, indicating that the result is set by ejecta mass and velocity rather than by high-$Y_e$ morphology. 
This is consistent with the weak morphology dependence reported by \citet{2021ApJ...918...10W}.

The epoch dependence of this constraining power is shown explicitly in Figure~\ref{fig:constraint_fraction}.
At the median monitoring depth of $J = 22.2$~mag scaled to $93$~Mpc, both TP and TS subsets reach a peak constraint fraction of $\sim\!0.97$ near $\sim\!1$~day post-merger and decline steeply beyond $\sim\!3$~days, falling effectively to zero by $\sim\!20$--30~days.
For both TP and TS morphologies, the constraint fraction remains above 50\% only over approximately $\sim\!0.25$--7~days after merger.
This demonstrates that NIR observations within this early window provide the primary constraining power against canonical KN models at this distance.
Consistent with this expectation, the strongest reductions in $S_{\rm cond}$ within our monitored sample are associated with host galaxies that received coverage within $\sim\!0.25$--7~days of the GW trigger, while hosts observed only at late epochs contribute substantially less to the overall constraint (Figure~\ref{fig:kn_constraint_example}).
While this benchmark is built on the canonical KN grid rather than on dedicated SSM models, the implication generalizes to canonical KN-like counterparts of BNS/NSBH mergers at comparable distances. 
Deep NIR observations of the highest-priority hosts obtained within $\sim\!0.25$--7~days after the trigger provide the most informative constraints against canonical KN-like emission at distances of order $\sim\!100$~Mpc, although observations before $\sim\!0.25$~days may have potential disadvantages.
Therefore, the resulting hybrid strategy, early observations for canonical KN emission followed by longer-baseline monitoring for super-KN and KN-in-SN, is better matched to constrain the range of plausible emission scenarios.

\subsection{Caveats and Host Prioritization}
\label{sec:caveats}
The host prioritization adopted here assumes that SSM merger rates trace stellar mass analogously to BNS and NSBH systems \citep{2019MNRAS.487....2M, 2019MNRAS.487.1675A, 2020MNRAS.495.1841A, 2020MNRAS.491.3419A}.
This assumption is theoretically unverified for SSM progenitors.
If progenitors of SSM form preferentially through channels less sensitive to stellar mass, $L_{\rm W1}$ may be a poor proxy for host probability, and $P_{\rm 3D}$-only weighting may be more appropriate.

As a representative sensitivity check, under $P_{\rm 3D}$-only weighting the same 59 monitored hosts cover 0.046\% of the cumulative prior, indicating that the monitored coverage fraction is itself sensitive to the choice of weighting scheme.
Indeed, if SSMs originate from channels that do not trace stellar mass, such as PBH, our fiducial coverage could be a significant overestimate. 
We therefore explicitly state that our current constraints are highly model-dependent regarding the progenitor-host association.
Consequently, a systematic sensitivity analysis across alternative weighting schemes, along with simulations constraining SSM host environments to calibrate host prioritization, is deferred to future work.
%


Independently of this uncertainty, the broad sky localization and finite queue time imposed a depth--coverage trade-off that limited the monitored sample to 59 hosts, meaning the source may simply reside outside the monitored subset.

\subsection{Implications for Future SSM Follow-up}
\label{sec:future}
The primary limitation of the present campaign is the depth--coverage trade-off imposed by the broad GW localization.
Our galaxy-targeted strategy constrains the counterpart parameter space effectively within the monitored subset but leaves 92.3\% of the weighted host prior unobserved.
Conversely, a strategy prioritizing wider area at shallower per-pointing depth would sample a larger fraction of the prior at the cost of detection sensitivity against fainter counterpart classes.
Neither approach can be optimally prescribed in advance for the SSM class: in the absence of established SSM counterpart light curves, both depth-limited and coverage-limited strategies retain blind spots, with the former missing event-wide counterparts and the latter missing intrinsically faint, delayed, or embedded ones.
The optimal balance therefore remains event- and model-dependent and is unlikely to be resolved until both dedicated counterpart models for SSM progenitors are better established and at least one SSM EM counterpart is actually observed.

Even with expanded coverage, the detection of super-KN or KN-in-SN counterparts requires monitoring baselines of order months rather than days.
Our campaign explicitly targets this regime but that lies well beyond the windows of typical optical follow-up campaigns.
In this context, archival revisitation of existing continuous all-sky survey data over the S251112cm localization area offers an immediately accessible complement to targeted follow-up.

Ongoing wide-field programs covering the whole sky periodically, including Pan-STARRS \citep{2016arXiv161205560C}, ATLAS \citep{2018PASP..130f4505T}, the All-Sky Automated Survey for SuperNovae (ASAS-SN; \citealt{2014ApJ...788...48S, 2017PASP..129j4502K}), and Rubin/LSST \citep{2019ApJ...873..111I}, have serendipitously accumulated multi-epoch optical coverage of the localization region spanning months to years.

Retrospective analysis of these archival datasets over a $\sim$\,months-long baseline may reveal slowly evolving transients that were below the detection threshold of individual rapid-cadence searches or that peaked after the initial follow-up window closed.
Similarly, SPHEREx, with its all-sky near-infrared spectrophotometric coverage \citep{2018arXiv180505489D}, provides an independent, repeated NIR dataset over the same sky regions that can help constrain redder, longer-lived counterpart classes without dedicated target-of-opportunity observations.
Within a single SPHEREx survey, an individual source is sampled across different wavelength channels over a days-to-weeks interval, while full-sky surveys are repeated on roughly six-month timescales. 
Consequently, systematic revisitation of these archives can extend the effective monitoring baseline to months, increasing the chance of discovering a delayed EM counterpart and strengthening constraints on SSM EM counterpart models, though actual detection of a transient still depends on chance alignment with SPHEREx's sampling cadence.

\section{Conclusion} \label{sec:conclusion}

We presented a two-month, galaxy-targeted UKIRT $J$-band follow-up campaign for the sub-solar-mass GW candidate S251112cm, monitoring 59 hosts at typical $5\sigma$ depths of $J \sim 22$--23~mag.
Consistent with all reported follow-up efforts for this event, we identify no convincing EM counterpart within the monitored host-galaxy subset.

At the inferred distance of $93$~Mpc, a Type~Ic-BL-like SN used as a proxy for the SN-like component of a KN-in-SN counterpart, an AT\,2017gfo-like KN, and delayed AT\,2017gfo-like KN motivated by delayed-merger scenarios would have remained detectable above our median depth at multiple epochs, disfavoring such counterparts within the monitored host subset.
The benchmark comparison with the canonical KN model grid shows that high-mass ejecta configurations are most strongly excluded, while faint configurations remain largely permissible.
The strongest constraints are driven by hosts observed within $\sim\!0.25$--7~days of the trigger, matching the epoch window over which more than 50\% of the canonical KN model grid is constrained.

These constraints are host-limited and conditional: the monitored sample encompasses only 7.7\% of the cumulative $P_{\rm 3D} \times L_{W1}$ weighted host prior, and the non-detection admits multiple physical interpretations, including a counterpart residing outside the monitored subset, an intrinsically fainter emission class such as a super-KN or KN-in-SN, a delay beyond our two-month baseline, or a non-EM progenitor channel such as a PBH merger.
Stellar-mass-based host weighting itself remains theoretically unverified for SSM progenitors.

In conclusion, our two-month UKIRT campaign provides deep NIR constraints on a prioritized subset of candidate S251112cm host galaxies, with no counterpart detected within the monitored sample.
Although host-limited, these observations translate the non-detection into empirical tests of counterpart scenarios that predict relatively NIR-bright emission on week-to-month timescales.
This work establishes a practical baseline for future SSM follow-up: deep, long-baseline NIR monitoring should be paired with improved localization, revisitation of overlapping all-sky survey archives to extend the search window, and constraints on the host-galaxy properties preferred by SSM progenitors.


\begin{acknowledgments}
G.S.H.P. acknowledges support from the Pan-STARRS project, which is a project of the Institute for Astronomy of the University of Hawai'i, and is supported by the NASA SSO Near Earth Observation Program under grants 80NSSC18K0971, NNX14AM74G, NNX12AR65G, NNX13AQ47G, NNX08AR22G, 80NSSC21K1572, and by the State of Hawai'i.

W.B.H. acknowledges support from the National Science Foundation Graduate Research Fellowship Program under Grant Nos. 1842402 and 2236415.
Any opinions, findings, conclusions, or recommendations expressed in this material are those of the author(s) and do not necessarily reflect the views of the National Science Foundation.

This work is based on observations obtained with the United Kingdom Infra-Red Telescope (UKIRT) under program 2025B program H03.
We sincerely thank the staff at UKIRT for their support and assistance in executing the Trigger-of-Opportunity observations.

\end{acknowledgments}




%
\facilities{UKIRT(WFCAM)}

\software{astropy \citep{2013A&A...558A..33A,2018AJ....156..123A,2022ApJ...935..167A},
          Source EXtractor \citep{1996A&AS..117..393B},
          SWarp \citep{2010ascl.soft10068B},
          HOTPANTS \citep{2015ascl.soft04004B}}

\appendix

\section{Monitored Galaxy Sample Catalog}
\label{appendix:galaxy_catalog}

This appendix presents the full list of 59 monitored host galaxies, ranked by the prioritization metric $P_{\rm 3D} \times L_{\rm W1}$ described in Section~\ref{sec:observation}.
Each Rank ID has the form G\#\#\#\#, where the numerical portion gives the galaxy's position in the full candidate-catalog ranking. Gaps therefore correspond to higher-ranked galaxies that were not monitored.

\startlongtable
  \begin{deluxetable*}{cccccc}
    \tablecaption{Monitored galaxy sample. The third column gives the three-dimensional localization probability density, $P_{3D}$, at each galaxy position and distance, scaled by $10^5$. The fourth column gives the host-priority weight, $P_{3D} \times L_{W1}$, scaled by $10^9$, where $L_{W1}$ is the WISE W1 luminosity used as a stellar-mass proxy.}
    \label{tab:galaxy}
    \tablehead{
      \colhead{Name} & \colhead{Rank ID} & \colhead{$P_{3D} / 10^5$} & \colhead{$P_{3D} \times L_{W1} / 10^{9}$} & \colhead{Distance (Mpc)} & \colhead{Rank}
      }
    \startdata
    IC 0564 & G0001 & 2.4 & 8.1 & 75.75 & 1 \\
    IC 1464B & G0002 & 0.83 & 7 & 105.88 & 2 \\
    WISEA J100012.04+093856.0 & G0003 & 2 & 6 & 78.82 & 3 \\
    UGC 05226 & G0004 & 2.2 & 5.4 & 66.86 & 4 \\
    MCG -01-24-010 & G0006 & 2.5 & 5 & 53.94 & 6 \\
    MRK 0421 & G0007 & 0.58 & 4.9 & 85.15 & 7 \\
    IC 1464A & G0008 & 0.76 & 4.8 & 107.09 & 8 \\
    NGC 7364 & G0009 & 0.67 & 4.8 & 76.48 & 9 \\
    IC 1471 & G0010 & 1.5 & 4.6 & 83.0 & 10 \\
    NGC 2951 NED01 & G0012 & 2.3 & 4.2 & 70.23 & 12 \\
    NGC 2951 NED02 & G0016 & 2.1 & 3.7 & 71.31 & 16 \\
    IC 0563 & G0018 & 2.5 & 3.6 & 57.0 & 18 \\
    UGCA 150 & G0020 & 0.48 & 3.5 & 28.38 & 20 \\
    NGC 7391 & G0021 & 0.49 & 3.4 & 54.75 & 21 \\
    NGC 2960 & G0022 & 0.72 & 3.4 & 76.97 & 22 \\
    NGC 2987 & G0042 & 1.3 & 2.5 & 54.46 & 42 \\
    IC 0561 & G0109 & 1.7 & 1.4 & 86.25 & 109 \\
    LSBC D709-06 & G0824 & 1.6 & 0.31 & 75.83 & 824 \\
    WISEA J110442.63+381405.6 & G0832 & 0.19 & 0.3 & 132.78 & 832 \\
    WISEA J110357.80+381416.3 & G0896 & 0.16 & 0.28 & 134.78 & 896 \\
    WISEA J230820.34-123541.6 & G1530 & 0.076 & 0.14 & 142.12 & 1530 \\
    WISEA J094619.25+025713.4 & G1569 & 2.7 & 0.14 & 59.16 & 1569 \\
    WISEA J110358.67+383604.8 & G1708 & 0.17 & 0.12 & 129.28 & 1708 \\
    WISEA J094602.22+030803.2 & G2066 & 1.4 & 0.087 & 89.28 & 2066 \\
    WISEA J230246.60-085930.6 & G2336 & 0.76 & 0.069 & 107.21 & 2336 \\
    WISEA J095956.25+093047.3 & G2682 & 2.1 & 0.051 & 77.7 & 2682 \\
    WISEA J230336.94-082732.6 & G2739 & 0.052 & 0.048 & 139.77 & 2739 \\
    WISEA J230302.33-083813.5 & G2817 & 0.055 & 0.046 & 141.33 & 2817 \\
    WISEA J100014.67+093702.1 & G2867 & 2.1 & 0.044 & 77.14 & 2867 \\
    WISEA J230639.62-123424.0 & G2917 & 0.047 & 0.042 & 146.16 & 2917 \\
    WISEA J110458.44+381545.3 & G2930 & 0.21 & 0.042 & 130.39 & 2930 \\
    WISEA J230303.20-082506.1 & G3104 & 0.71 & 0.036 & 105.75 & 3104 \\
    WISEA J110409.98+381203.6 & G3243 & 0.18 & 0.032 & 133.4 & 3243 \\
    WISEA J110450.89+383621.2 & G3394 & 0.17 & 0.028 & 129.96 & 3394 \\
    WISEA J224446.85-000513.8 & G3550 & 0.57 & 0.025 & 72.74 & 3550 \\
    WISEA J110207.20+383851.8 & G3703 & 0.1 & 0.021 & 123.99 & 3703 \\
    WISEA J094618.06+032500.0 & G3712 & 2 & 0.021 & 81.54 & 3712 \\
    WISEA J110420.40+381155.1 & G3717 & 0.2 & 0.021 & 131.61 & 3717 \\
    WISEA J110246.81+381336.4 & G4281 & 0.2 & 0.013 & 125.26 & 4281 \\
    WISEA J230833.02-123639.9 & G4461 & 0.069 & 0.011 & 143.47 & 4461 \\
    WISEA J110415.43+381416.7 & G4896 & 0.22 & 0.0071 & 129.63 & 4896 \\
    IC 0549 & G5066 & 0.19 & 0.006 & 22.7 & 5066 \\
    WISEA J110420.08+381239.8 & G5237 & 0.21 & 0.0048 & 130.93 & 5237 \\
    2MASS J09462238+0304220 & G6577 & 1.7 & 0 & 86.94 & 6577 \\
    2MASS J09461985+0304081 & G6602 & 2.5 & 0 & 73.61 & 6602 \\
    2MASS J11042824+3812401 & G6662 & 0.13 & 0 & 138.0 & 6662 \\
    2MASS J23025897-0826154 & G6671 & 0.68 & 0 & 106.82 & 6671 \\
    ARP 303 & G6315 & 1.4 & 0 & 89.55 & 6315 \\
    {[HB89] 1101+384 ABS01} & G8873 & 0.15 & 0 & 44.11 & 8873 \\
    USGC U256 & G8787 & 1.6 & 0 & 87.55 & 8787 \\
    6dF J0911182-082839 & G8742 & 0.044 & 0 & 99.06 & 8742 \\
    LEDA 4677511 & G8708 & 0.79 & 0 & 106.58 & 8708 \\
    IC 1464 & G8510 & 0.76 & 0 & 107.74 & 8510 \\
    MCG +07-23-014 & G8025 & 0.26 & 0 & 112.5 & 8025 \\
    {Mr18:[BFW2006] 02400} & G7951 & 0.047 & 0 & 140.54 & 7951 \\
    {Mr18:[BFW2006] 05582} & G7278 & 0.79 & 0 & 106.58 & 7278 \\
    SDSS J094053.61+033903.6 & G6908 & 0.86 & 0 & 52.66 & 6908 \\
    SDSS J224424.47-000953.5 & G6698 & 0.71 & 0 & 70.41 & 6698 \\
    SDSS J110458.85+381023.5 & G6697 & 0.11 & 0 & 145.28 & 6697 \\
    \enddata
  \end{deluxetable*}

\section{Alternative Heatmap Representations and Diagnostics}
\label{appendix:alt_heatmaps}

\begin{figure*}[t]
  \centering
  \includegraphics[width=1.0\textwidth]{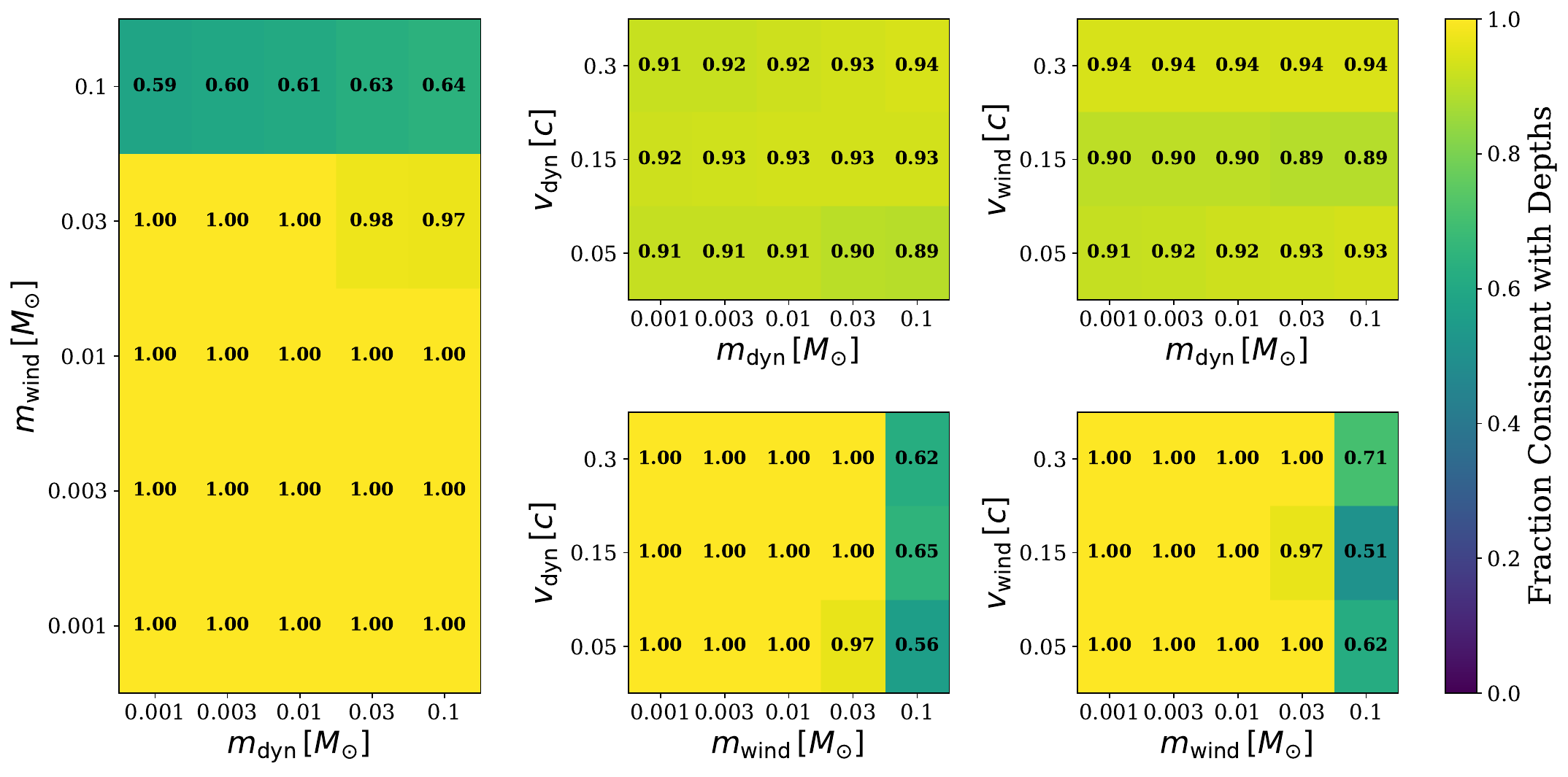}
  \caption{Legacy OR-gate diagnostic heatmap for the full monitored subset in the TP representation. This map does not represent the conditional weighted consistent fraction used in Figure~\ref{fig:global_tp_scond}.}
  \label{fig:appendix_tp_all}
\end{figure*}

\begin{figure*}[t]
  \centering
  \includegraphics[width=1.0\textwidth]{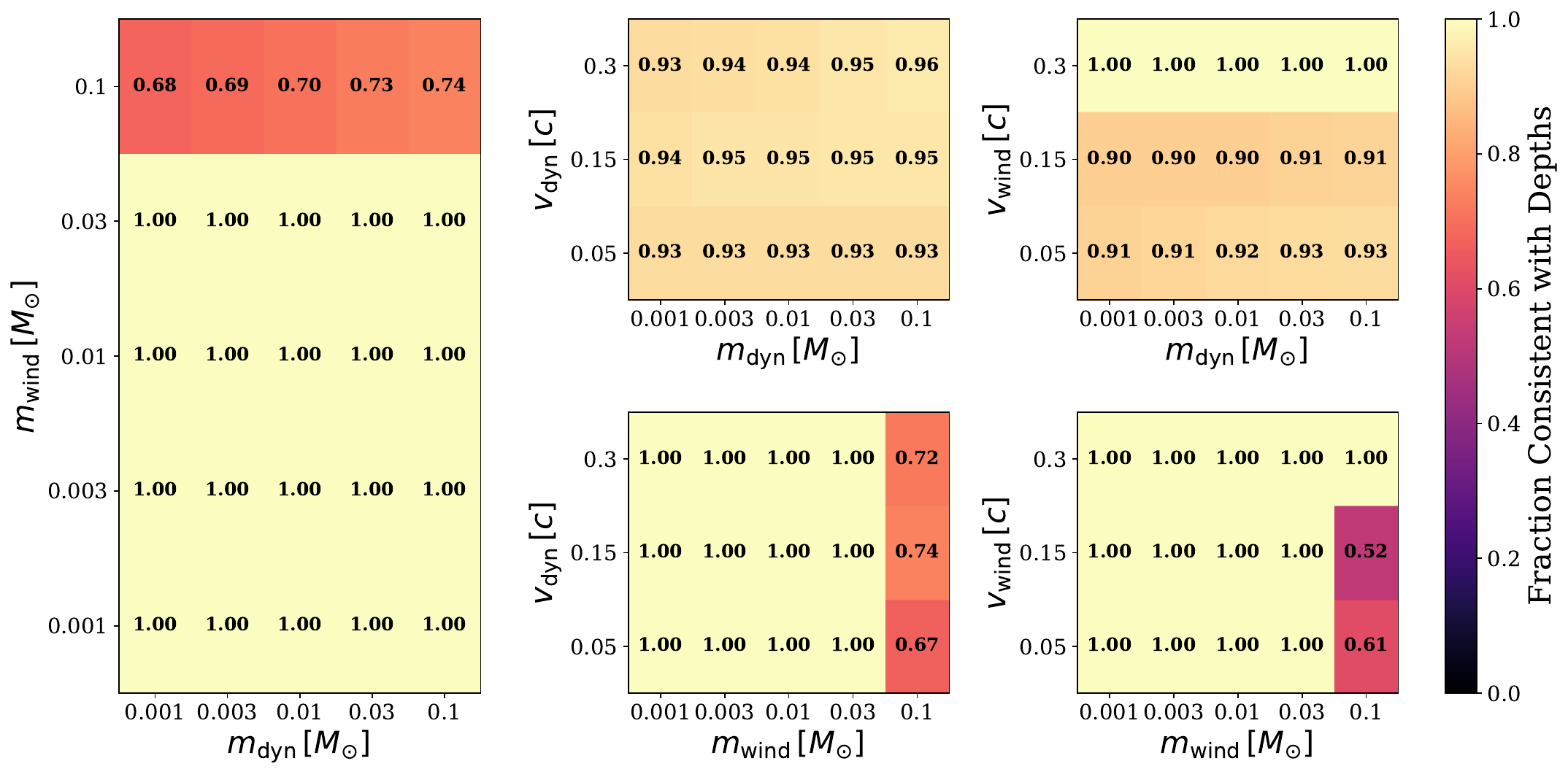}
  \caption{Legacy OR-gate diagnostic heatmap for the full monitored subset in the TS representation. As in Figure~\ref{fig:appendix_tp_all}, this view does not represent the conditional weighted consistent fraction used in Figure~\ref{fig:global_ts_scond}.}
  \label{fig:appendix_ts_all}
\end{figure*}

For completeness, Figures~\ref{fig:appendix_tp_all} and \ref{fig:appendix_ts_all} show legacy OR-gate heatmaps for the same benchmark model grid.
Unlike the conditional weighted consistent-fraction maps in Section~\ref{subsec:condition}, these diagnostics count a model as consistent if it remains fainter than the observational limits for at least one monitored host.


\bibliography{sample701}{}
\bibliographystyle{aasjournalv7}



\end{document}